\documentclass[aps,showpacs,preprint,amsmath,floatfix]{revtex4}

\usepackage{graphicx}
\usepackage{epsfig}
\usepackage{dcolumn}
\usepackage{bm}

\begin{document}

\def\bea{\begin{eqnarray}}
\def\eea{\end{eqnarray}}
\def\be{\begin{equation}}
\def\ee{\end{equation}}

\title{Recursive parametrization of Quark flavour mixing
matrices}

\author{S.~Chaturvedi}

\email[]{scsp@uohyd.ernet.in}

\affiliation{
School of Physics.\\
University of Hyderabad\\
Hyderabad 500 046, India.
}

\author{V.~Gupta}

\email[]{virendra@mda.cinvestav.mx}

\author{G.~S\'anchez-Col\'on}

\email[]{gsanchez@mda.cinvestav.mx}

\affiliation{
Departamento de F\'{\i}sica Aplicada.\\
Centro de Investigaci\'on y de Estudios Avanzados\\
del Instituto Polit\'ecnico Nacional.\\
Unidad M\'erida.\\
A.P. 73, Cordemex.\\
M\'erida, Yucat\'an, 97310. M\'EXICO.
}

\author{N. Mukunda}

\email[]{nmukunda@cts.iisc.ernet.in}

\affiliation{
Centre for High Energy Physics,\\
Indian Institute of Science.\\
Bangalore 560012, India
}

\date{\today}

\begin{abstract}

We examine quark flavour mixing matrices for three and four generations using
the recursive parametrization of $U(n)$ and $SU(n)$ matrices developed by some
of us in Refs.~\cite{2} and \cite{3}. After a brief summary of the recursive
parametrization, we obtain expressions for the independent rephasing invariants
and also the constraints on them that arise from the requirement of mod symmetry
of the flavour mixing matrix.

\end{abstract}


\maketitle

\section{\label{introduction}Introduction}

Study of flavor mixing in weak interactions provides a low energy
window for new physics. Currently, experiments are underway at
Belle and BaBar to check the \lq\lq unitarity triangle\rq\rq for
the $3\times 3$ flavor mixing matrix, as accurately as possible.
If there is a significant deviation then it would be a signal for
the existence of more than three generations. Furthermore, the
$3\times 3$ CKM mixing matrix contains only one CP-violating
phase thus implying that CP-violations in different processes
are related. Again, the violation of any one of these relations
would be a signal for more generations. Consequently, in this
paper we study some general  properties of a $4\times 4$ flavor
mixing matrix. Such a matrix in general has six angles and
three phases. However, a moduli symmetric $4\times 4$ unitary
matrix has fewer parameters. We study such a matrix in detail
and present parametrizations which would be useful for
confrontation with experiments in the future.

In Sec.~\ref{secdos}, rephasing invariants for a $n\times n$
unitary matrices are defined. In addition, relations between
plaquettes for the particular cases $n=3$ and 4 are given. In
Sec.~\ref{sectres}, recursive parametrization for the $n\times n$
case is given together with that for $n=2$, 3, and 4.
Rephasing invariants in the recursive parametrization are
presented in Sec.~\ref{seccuatro}. A moduli symmetric unitary
matrix has fewer parameters and the results for $n=3$ and 4 are
given in Sec.~\ref{seccinco}. In Sec.~\ref{secseis} the standard PDG
parametrization~\cite{pdg08} is obtained using the recursive
approach. The conclusions are presented in Sec.~\ref{conclusions}.

\section{\label{secdos}Rephasing invariants of ${\bm U(n)}$
matrices}

It is known that for a given $U(n)$ matrix ($n\times n$ unitary
matrix) $V$ under rephasing, {\it i.e.}, under multiplication
by independent diagonal $U(n)$ matrices from the left and the
right,

\be
V\rightarrow V^\prime = D(\theta^\prime)VD(\theta),
\ee

\noindent
with

\be
D(\theta)={\rm diag}\{e^{i\theta_1},\ldots,e^{i\theta_n}\}
\quad
{\rm and}
\quad
D(\theta^\prime)={\rm diag}\{e^{i{\theta_1}^\prime},\ldots,e^{i\theta_n^\prime}\},
\ee

\noindent
the basic quantities that remain invariant are (a) the
$n^2$ moduli $|V_{ij}|$ ($i,j=1,\ldots,n$) and (b) the
$n^2(n-1)^2/4$ Bargmann invariants or \lq\lq plaquettes\rq\rq,
$\Delta_{ijkl}\equiv V_{ik}V_{jk}^*V_{jl}V_{il}^*$ ($i<j$,
$k<l$; $i,j,k,l=1,\ldots,n$). Unitarity of $V$ gives rise to
relations between them and one finds that there are $n(n-1)/2$
independent invariants of type (a), {\it i.e.}, of modulus-type.
As to those of type (b), the phase-type invariants, using
unitarity one finds that the $n^2(n-1)^2/4$ invariants can all
be algebraically expressed in terms of $(n-1)^2$ elementary
plaquettes $\Delta_{ik}\equiv \Delta_{i,i+1,k,k+1}$. The
$(n-1)^2$ basic plaquettes, in turn, are related to each other
and can be expressed in terms of $(n-1)(n-2)/2$ algebraically
independent primitives which may be taken to be $\Delta_{ik}$ ($i
< k\leq n-1$). The total number of algebraically independent
rephasing invariants is thus $n(n-1)/2 +(n-1)(n-2)/2$, {\it
i.e.}, $(n-1)^2$. For $n=3$ and $n=4$, the relations are
explicitly given below.

\subsection{\label{subsecdosa}Relations between plaquettes for
${\bm n=3}$}

Here there is only one primitive, {\it viz.}, $\Delta_{12}$.
Orthogonality of the rows of $V$ gives the relations,

\be
\Delta_{i1}=-|V_{i2}|^2|V_{i+12}|^2-\Delta_{i2}.
\ee

\noindent
Likewise the orthogonality of the columns gives,

\be
\Delta_{1i}=-|V_{2i}|^2|V_{2i+1}|^2-\Delta_{2i} .
\ee

These are four inhomogeneous equations for four quantities
$\Delta_{11}$, $\Delta_{21}$, $\Delta_{22}$, and $\Delta_{12}$.
One of them is derivable from the other three leaving us with
three equations which allow us to solve for
$\Delta_{11}$, $\Delta_{21}$, and $\Delta_{22}$ in terms of
$\Delta_{12}$:

\bea
\Delta_{11}&=&-|V_{12}|^2|V_{22}|^2-\Delta_{12}^*,\nonumber\\
\Delta_{22}&=& -|V_{22}|^2|V_{23}|^2-\Delta_{12}^*,\\
\Delta_{21}&=&|V_{22}|^2(|V_{23}|^2-|V_{32}|^2)+\Delta_{12}.\nonumber
\eea

\noindent
Any other plaquette, {\it e.g.}, $\Delta_{1213}$, can be
expressed as $\Delta_{11}\Delta_{12}/|V_{12}|^2|V_{22}|^2$ and,
using the relations above, as
$-|V_{13}|^2|V_{23}|^2-\Delta_{12}$.

As is well known, these relations have the consequence that the
imaginary parts of all the plaquettes are the same, up to a sign.
Furthermore, if even one $V_{ij}$, say $V_{11}$, vanishes, then
all the plaquettes become real. It is also evident that,
imposing mod symmetry on $V$, {\it i.e.}, requiring
$|V_{ij}|=|V_{ji}|$, while reducing the number of independent
modulus type invariants from three to two, has no effect on the
number of independent phase type invariants.

\subsection{\label{subsecdosb}Relations between plaquettes for
${\bm n=4}$}

In this case row and column orthogonality of $V$ respectively
yield:

\be
|V_{i3}|^2|V_{i+13}|^2(\Delta_{i1}+\Delta_{i2}^*)
=-\Delta_{i2}^*(\Delta_{i2}+\Delta_{i3}^*)\quad (i=1,2,3),
\label{cinco}
\ee

\be
|V_{3i}|^2|V_{3i+1}|^2(\Delta_{1i}+\Delta_{2i}^*)
=-\Delta_{2i}^*(\Delta_{2i}+\Delta_{3i}^*)\quad (i=1,2,3),
\label{sies}
\ee

\noindent
which may alternatively be written as

\be
|V_{i2}|^2|V_{i+12}|^2(\Delta_{i3}+\Delta_{i2}^*)
=-\Delta_{i2}^*(\Delta_{i2}+\Delta_{i1}^*)\quad (i=1,2,3),
\label{siete}
\ee

\be
|V_{2i}|^2|V_{2i+1}|^2(\Delta_{3i}+\Delta_{2i}^*)
=-\Delta_{2i}^*(\Delta_{2i}+\Delta_{1i}^*)\quad (i=1,2,3).
\label{ocho}
\ee

Choosing Eqs.~(\ref{cinco}) and (\ref{ocho}) we have:

\bea
|V_{13}|^2|V_{23}|^2(\Delta_{11}+\Delta_{12}^*)
&=&-\Delta_{12}^*(\Delta_{12}+\Delta_{13}^*),\nonumber\\
|V_{23}|^2|V_{33}|^2(\Delta_{21}+\Delta_{22}^*)
&=&-\Delta_{22}^*(\Delta_{22}+\Delta_{23}^*),\nonumber\\
|V_{33}|^2|V_{43}|^2(\Delta_{31}+\Delta_{32}^*)
&=&-\Delta_{32}^*(\Delta_{32}+\Delta_{33}^*),\\
|V_{21}|^2|V_{22}|^2(\Delta_{31}+\Delta_{21}^*)
&=&-\Delta_{21}^*(\Delta_{21}+\Delta_{11}^*),\nonumber\\
|V_{22}|^2|V_{23}|^2(\Delta_{32}+\Delta_{22}^*)
&=&-\Delta_{22}^*(\Delta_{22}+\Delta_{12}^*),\nonumber\\
|V_{23}|^2|V_{24}|^2(\Delta_{33}+\Delta_{23}^*)
&=&-\Delta_{23}^*(\Delta_{23}+\Delta_{13}^*).\nonumber
\eea

\noindent
These six equations for the nine plaquettes allow us to solve
all of them in terms of the primitives which we choose to be
$\Delta_{12}$, $\Delta_{13}$, and $\Delta_{23}$. The relevant
equations are:

\be
\Delta_{11}=-|V_{12}|^2|V_{22}|^2-\Delta_{12}^*
\left(1+\frac{\Delta_{13}^*}{|V_{13}|^2|V_{23}|^2}\right)
\ee

\be
\Delta_{33}=-|V_{33}|^2|V_{34}|^2-\Delta_{23}^*
\left(1+\frac{\Delta_{13}^*}{|V_{23}|^2|V_{24}|^2}\right)
\ee

\be
\left(1+\frac{\Delta_{11}}{|V_{21}|^2|V_{22}|^2}\right)\Delta_{21}
- \left(1+ \frac{\Delta_{33}}{|V_{33}|^2|V_{43}|^2}\right)
\Delta_{32} = |V_{32}|^2(|V_{42}|^2-|V_{31}|^2)
\ee

\bea
\lefteqn{\left(1+\frac{\Delta_{12}^*}{|V_{22}|^2|V_{23}|^2}\right)\Delta_{21}
- \left(1+\frac{\Delta_{23}^*}{|V_{23}|^2|V_{33}|^2}\right)
\Delta_{32} =} \nonumber\\
&&|V_{32}|^2|V_{33}|^2\left(1+\frac{\Delta_{23}^*}
{|V_{23}|^2|V_{33}|^2}\right) - |V_{22}|^2|V_{32}|^2
\left(1+\frac{\Delta_{12}^*}{|V_{22}|^2|V_{23}|^2}\right) \eea

\be
\Delta_{31}
= -|V_{31}|^2|V_{32}|^2-\Delta_{21}^*\left(1+\frac{\Delta_{11}^*}
{|V_{21}|^2|V_{22}|^2}\right)
\ee

\be
\Delta_{22} =
\frac{\Delta_{21}^*+|V_{22}|^2|V_{32}|^2}{\left(1
+ \displaystyle{\frac{\Delta_{23}}{|V_{23}|^2|V_{33}|^2}}\right)}.
\ee

\section{\label{sectres}Recursive parametrization of ${\bm U(n)}$
(${\bm SU(n)}$) matrices}

Let $U(n)$ denote the group of unitary matrices acting on all $n$
dimensions. For $m=1,2,\ldots ,n-1$, we will denote by $U(m)$ the
unitary group acting on the first $m$ dimensions, leaving the
dimensions $m+1,\ldots,n$, unaffected. Then we have the canonical
subgroup chain

\begin{equation}
U(1)\subset U(2)\subset\cdots \subset U(n-1)\subset U(n).
\end{equation}

\noindent
General matrices of $U(n),U(n-1),\ldots$ will be written as
${\mathcal A}_n,{\mathcal A}_{n-1},\ldots$, respectively. In a matrix
${\mathcal A}_m\in U(m)$ the last rows and columns are trivial,
with ones along the diagonals and zeros elsewhere (when no
confusion is likely to arise, ${\mathcal A}_m$ will also denote
an unbordered $m\times m$ unitary matrix).

It was shown in~\cite{2} that any matrix ${\mathcal A}_n\in
U(n)$ can be expressed uniquely as an $n$-fold product

\begin{equation}
{\mathcal A}_n = {\mathcal A}_n({\bm\zeta})
{\mathcal A}_{n-1}({\bm\eta})
{\mathcal A}_{n-2}({\bm\xi}) \cdots
{\mathcal A}_4({\bm\gamma})
{\mathcal A}_3({\bm\beta})
{\mathcal A}_2({\bm\alpha})
{\mathcal A}_1(\chi),
\end{equation}

\noindent
where ${\mathcal A}_n({\bm\zeta})$ is a special $U(n)$ element
determined by an $n$-component complex unit vector
${\bm\zeta}$, ${\mathcal A}_{n-1}({\bm\eta})$ is a special
$U(n-1)$ element determined by an $n-1$-component complex unit
vector ${\bm\eta}$, and so on down to ${\mathcal
A}_2({\bm\alpha})$ that is a special $U(2)$ element determined
by a two-component complex unit vector ${\bm\alpha}$, and
${\mathcal A}_1(\chi)$ is a phase factor belonging to $U(1)$.
The complex unit vectors $\{{\bm\zeta}, {\bm\eta}, \ldots\}$,
appear as the last columns of the (unbordered) matrices
$\{{\mathcal A}_n({\bm\zeta}), {\mathcal A}_{n-1}({\bm\eta}),
\ldots\}$ and can be identified with the labels of the cosets
$\{U(n)/U(n-1), U(n-1)/U(n-2),\ldots\}$.  Remembering that
$\{{\bm\zeta}, {\bm\eta}, \ldots\}$ are complex unit vectors of
dimensions $\{n, n-1, \ldots\}$, it is easily seen that the number
of real independent parameters add up to $n^2$ as they should.

The same considerations as above apply to $SU(n)$ matrices as well.
Denoting by $A_n({\bm\zeta})$ the corresponding matrices in
$SU(n)$, any $A_n \in SU(n)$ can be decomposed as

\begin{equation}
A_n= A_n({\bm\zeta})
A_{n-1}({\bm\eta})
A_{n-2}({\bm\xi})\cdots
A_4({\bm\gamma})
A_3({\bm\beta})
A_2({\bm\alpha})
\end{equation}

The above construction fixes only the last column of the unitary
matrix ${\mathcal A}_n({\bm\zeta})$ as ${\bm\zeta}$, and one has
a great deal of freedom in arranging the remaining $n-1$ columns
leading to many explicit forms for these matrices. In this work
we consider two explicit forms which correspond to those
discussed in~\cite{2} and \cite{3} respectively.

The explicit expressions for the nonzero matrix elements of
${\mathcal A}_n({\bm\zeta})$ considered in~\cite{2} are

\bea
{\mathcal A}_n({\bm\zeta})&=& (a_{jk}({\bm\zeta}))\in U(n);
\quad j,k=1,2,\ldots,n.\nonumber\\
a_{jn}({\bm\zeta})&=&\zeta_j;
\quad j=1,2,\ldots,n.\\
a_{jj-1}({\bm\zeta})&=&\rho_{j-1}/\rho_{j};
\quad j=2,3,\ldots,n;\nonumber\\
&&
\rho_j=\sqrt{1-|\zeta_{j+1}|^2-|\zeta_{j+2}|^2-\cdots-|\zeta_n|^2}=
\sqrt{|\zeta_{1}|^2+\cdots +|\zeta_j|^2}.\nonumber\\
a_{jk}({\bm\zeta})&=&-\zeta_j\zeta_{k+1}^*/\rho_{k}\rho_{k+1};
\quad j\leq k\leq n-1.\nonumber
\eea

\noindent
Thus, for instance, for $n=2,3,4$ we have:

\be
{\mathcal A}_2({\bm\alpha})=
\left(
\begin{array}{cc}
-\alpha_2^*\alpha_1/\mu_1&\alpha_1\\
\mu_1&\alpha_2
\end{array}
\right),\qquad \mu_1=|\alpha_1|.
\ee

\be
{\mathcal A}_3({\bm\beta})=
\left(
\begin{array}{ccc}
-\beta_2^*\beta_1/\sigma_1\sigma_2&-\beta_3^*\beta_1/\sigma_2&\beta_1\\
\sigma_1/\sigma_2&-\beta_3^*\beta_2/\sigma_2&\beta_2\\
0&\sigma_2&\beta_3
\end{array}
\right),\qquad \sigma_1=|\beta_1|,\quad
\sigma_2=\sqrt{|\beta_1|^2+|\beta_2|^2}.
\ee

\bea
{\mathcal A}_4({\bm\gamma})&=&
\left(
\begin{array}{cccc}
-\gamma_2^*\gamma_1/\rho_1\rho_2&-\gamma_3^*\gamma_1/\rho_2\rho_3&
-\gamma_4^*\gamma_1/\rho_3&\gamma_1\\
\rho_1/\rho_2&-\gamma_3^*\gamma_2/\rho_2\rho_3&
-\gamma_4^*\gamma_2/\rho_3&\gamma_2\\
0&\rho_2/\rho_3&-\gamma_4^*\gamma_3/\rho_3&\gamma_3\\
0&0&\rho_3&\gamma_4
\end{array}
\right),\nonumber\\
&&\rho_1=|\gamma_1|,\quad\rho_2=\sqrt{|\gamma_1|^2+\gamma_2|^2},
\quad\rho_3=\sqrt{|\gamma_1|+|\gamma_2|+|\gamma_3|^2}.
\eea

The determinant of the matrices ${\mathcal A}_n({\bm\zeta})$
turns out to be $(-1)^{n-1}\zeta_1/|\zeta_1|$ and hence the
corresponding $SU(n)$ matrices can be obtained by multiplying,
for instance, the first column by
$(-1)^{n-1}\zeta_1^*/|\zeta_1|$. Thus for $n=2,3,4$ we have

\be
 A_2({\bm\alpha})=
\left(
\begin{array}{cc}
\alpha_2^*&\alpha_1\\
-\alpha_1^*&\alpha_2
\end{array}
\right),
\ee

\be
A_3({\bm\beta})=
\left(
\begin{array}{ccc}
-\beta_2^*/\sigma_2&-\beta_3^*\beta_1/\sigma_2&\beta_1\\
\beta_1^*/\sigma_2& -\beta_3^*\beta_2/\sigma_2&\beta_2\\
0&\sigma_2&\beta_3
\end{array}
\right),
\ee

\be
A_4({\bm\gamma})=
\left(\begin{array}{cccc}
\gamma_2^*/\rho_2&-\gamma_3^*\gamma_1/\rho_2\rho_3&
-\gamma_4^*\gamma_1/\rho_3&\gamma_1\\
-\gamma_1^*/\rho_2&-\gamma_3^*\gamma_2/\rho_2\rho_3&
-\gamma_4^*\gamma_2/\rho_3&\gamma_2\\
0&\rho_2/\rho_3&-\gamma_4^*\gamma_3/\rho_3&\gamma_3\\
0&0&\rho_3&\gamma_4
\end{array}
\right).
\ee

\noindent
This parametrization assumes that $\zeta_1$ is nonzero. As a
result, in the extreme case when ${\bm\zeta}=(1,0,\ldots,0)$, the
matrix ${\mathcal A}_n({\bm\zeta})$ does not reduce to the
identity matrix. A parametrization where this does happen and
which corresponds to that given in~\cite{3} is given below:

\bea
{\mathcal A}_n({\bm\zeta})&=& (a_{jk}({\bm\zeta}))\in U(n);
\quad j,k=1,2,\ldots,n.\nonumber\\
a_{jn}({\bm\zeta})&=& \zeta_j;
\quad j=1,2,\ldots,n.\\
a_{jj}({\bm\zeta})&=& \rho_j/\rho_{j-1};
\quad j=1,2,\ldots,n-1;
\quad\rho_0=1;\nonumber\\
&&
\rho_j=\sqrt{1-|\zeta_1|^2-|\zeta_2|^2-\cdots-|\zeta_j|^2}=
\sqrt{|\zeta_{j+1}|^2+\cdots|\zeta_n|^2}.\nonumber\\
a_{jk}({\bm\zeta})&=& -\zeta_j\zeta_k^*/(\rho_{k-1}\rho_{k});
\quad j>k,\quad k=1,2,\ldots,n-1.\nonumber
\eea

\noindent
Note that we are using the same symbols as in the parametrization
earlier though with different meanings. For $n=2,3,4$, we
explicitly have:

\be
{\mathcal A}_2({\bm \alpha})=
\left(
\begin{array}{cc}
\mu_1&\alpha_1\\
-\alpha_1^*\alpha_2/\mu_1&\alpha_2
\end{array}
\right);
\quad\mu_1=|\alpha_2|.
\ee

\be
{\mathcal
A}_3({\bm\beta})=
\left(
\begin{array}{ccc}
\sigma_1&0&\beta_1\\
-\beta_1^*\beta_2/\sigma_1&\sigma_2/\sigma_1&\beta_2\\
-\beta_1^*\beta_3/\sigma_1&-\beta_2^*\beta_3/\sigma_1\sigma_2&\beta_3
\end{array}
\right);
\quad\sigma_1=\sqrt{|\beta_2|^2+|\beta_3|^2},
\quad\sigma_2=|\beta_3|.
\ee

\bea
{\mathcal A}_4({\bm\gamma})&=&
\left(
\begin{array}{cccc}
\rho_1&0&0&\gamma_1\\
-\gamma_1^*\gamma_2/\rho_1&\rho_2/\rho_1&0 &\gamma_2\\
-\gamma_1^*\gamma_3/\rho_1&-\gamma_2^*\gamma_3/\rho_1\rho_2&
\rho_3/\rho_2&\gamma_3\\
-\gamma_1^*\gamma_4/\rho_1&-\gamma_2^*\gamma_4/\rho_1\rho_2&
-\gamma_3^*\gamma_4/\rho_2\rho_3&\gamma_4
\end{array}
\right);
\nonumber\\
&&
\rho_1=\sqrt{|\gamma_2|^2+|\gamma_3|^2+|\gamma_4|^2},
\quad
\rho_2=\sqrt{|\gamma_3|^2+|\gamma_4|^2},
\quad
\rho_3=|\gamma_4|.
\eea

\noindent
The determinant of the matrices ${\mathcal A}_n({\bm\zeta})$ is
$\zeta_n/|\zeta_n|$. We can convert the above matrices to $SU(n)$
matrices by multiplying, say the $(n-1)$-th column by
$\zeta_n^*/|\zeta_n|$. Thus, for $n=2,3,4$, we have:

\be
A_2({\bm\alpha})=
\left(\begin{array}{cc}
\alpha_2^*&\alpha_1\\
-\alpha_1^*&\alpha_2
\end{array}
\right),
\label{35}
\ee

\be
A_3({\bm\beta})=
\left(\begin{array}{ccc}
\sigma_1&0&\beta_1\\
-\beta_1^*\beta_2/\sigma_1&\beta_3^*/\sigma_1&\beta_2\\
-\beta_1^*\beta_3/\sigma_1&-\beta_2^*/\sigma_1&\beta_3\\
\end{array}
\right),
\label{36}
\ee

\be
A_4({\bm\gamma})=
\left(
\begin{array}{cccc}
\rho_1&0&0&\gamma_1\\
-\gamma_1^*\gamma_2/\rho_1&\rho_2/\rho_1&0 &\gamma_2\\
-\gamma_1^*\gamma_3/\rho_1&-\gamma_2^*\gamma_3/\rho_1\rho_2&
\gamma_4^*/\rho_2&\gamma_3\\
-\gamma_1^*\gamma_4/\rho_1&-\gamma_2^*\gamma_4/\rho_1\rho_2&
-\gamma_3^*/\rho_2&\gamma_4
\end{array}
\right).
\label{37}
\ee

Given a matrix ${\mathcal A}_n \in U(n)$, we can determine the
parameters, the complex unit vectors,
$\{{\bm\zeta},{\bm\eta},\ldots\}$ in a recursive fashion through
the following steps.

\begin{itemize}
\item
Write ${\mathcal A}_n=(a_{jk})\in U(n)$ as

\begin{equation}
{\mathcal A}_n={\mathcal A}_n({\bm\zeta}){\mathcal B}_{n-1},
\end{equation}

where ${\bm\zeta}$ is the last column of ${\mathcal A}_n$

\begin{equation}
\zeta_j=a_{jn}.
\end{equation}

\item
With ${\mathcal A}_n({\bm\zeta})$ thus determined, we have

\begin{equation}
{\mathcal B}_{n-1}={\mathcal A}_n^\dagger({\bm\zeta}){\mathcal
A}_n.
\end{equation}

The matrix elements $(b_{ij})$, $n-1\geq i,j\geq 1$, of
${\mathcal B}_{n-1}$ in the first form~\cite{2} are given by

\be
b_{ij}=
\sum_{k=1}^{n}a_{ki}^*({\bm\zeta})a_{kj}
=-\frac{1}{\rho_i\rho_{i+1}}
\sum_{k=i+1}^{n}a_{kn}^*(a_{kj}a_{i+1n}-a_{kn}a_{i+1j})
\ee

and in the second form~\cite{3} by

\be
b_{ij}=
\sum_{k=1}^{n}a_{ki}^*({\bm\zeta})a_{kj}=
\frac{1}{\rho_i\rho_{i-1}}
\sum_{k=i+1}^{n}a_{kn}^*(a_{ij}a_{kn}-a_{in}a_{kj}).
\ee

\item
Write ${\mathcal B}_{n-1}$ as

\begin{equation}
{\mathcal B}_{n-1}={\mathcal
A}_{n-1}({\bm\eta}){\mathcal C}_{n-2}
\end{equation}

with

\begin{equation}
\eta_j=b_{jn-1}.
\end{equation}

\item
Repeat the same procedure as above with ${\mathcal C}_{n-2}$.
\end{itemize}

The same procedure applies to the decomposition of an $SU(n)$
matrix. Thus, for instance, using the second form~\cite{3}, a matrix
$V\in SU(3)$ can be decomposed as

\begin{equation}
V=A_3({\bm\beta})A_2({\bm\alpha}),
\end{equation}

\noindent
where

\be
\beta_1=V_{13},\quad\beta_{2}=V_{23},\quad\beta_3=V_{33},
\ee

\noindent
and

\be
\alpha_1=
\frac{[V_{23}^*(V_{12}V_{23}-V_{13}V_{22})+
V_{33}^*(V_{12}V_{33}-V_{13}V_{32})]}{{\sqrt{|V_{23}|^2+|V_{33}|^2}}},
\quad
\alpha_2=
\frac{(V_{33}V_{22}-V_{32}V_{23})}{\sqrt{|V_{23}|^2+|V_{33}|^2}}.
\ee

\section{\label{seccuatro}Rephasing invariants in the recursive parametrization}

Having shown how to parametrize a given $U(n)$ ($SU(n)$) matrix in terms of a
sequence of complex unit vectors $\{{\bm\zeta},{\bm\eta},\ldots\}$ of
dimensions $\{n, n-1, \ldots\}$, we now examine how these parameters transform
under rephasing with the purpose of constructing rephasing invariants out of
them. For simplicity and without any loss of generality we will assume that the
given matrix $V$ belongs to $SU(n)$ and will consider the cases $n=3,4$ and
discuss the transformation properties of the parameters in both the
forms~\cite{2,3} given above. In the first form~\cite{2}, any $SU(3)$ can be
written as

\bea
V&=&
A_3({\bm\beta})A_2({\bm\alpha})\nonumber\\
&=&
\left(
\begin{array}{ccc}
-\beta_2^*\alpha_2^*/\sigma_2 + \beta_3^*\beta_1\alpha_1^*/\sigma_2&
-\beta_2^*\alpha_1 - \beta_3^*\beta_1\alpha_2\sigma_2 & \beta_1\\
\beta_1^*\alpha_2^*/\sigma_2 + \beta_3^*\beta_2\alpha_1^*/\sigma_2
& \beta_1^*\alpha_1/\sigma_2 - \beta_3^*\beta_2\alpha_2/\sigma_2 & \beta_2\\
-\sigma_2\alpha_1^* & \sigma_2\alpha_2 & \beta_3
\end{array}
\right).
\label{49}
\eea

\noindent
Under rephasing by independent diagonal $SU(3)$ matrices $D(\theta)$ and
$D(\theta^\prime)$ where $D(\theta)={\rm
diag}(e^{i(\theta_1+\theta_2)},e^{i(-\theta_1+\theta_2)},e^{-2i\theta_2})$
and $D(\theta^\prime)$ is similarly defined, we have

\bea
V\rightarrow V^\prime&=&D(\theta^\prime) V D(\theta)\nonumber\\
&=& A_3({\bm\beta'}) A_2({\bm\alpha'})
\eea

\noindent
From the locations of $\alpha_1$, $\alpha_2$, $\beta_1$, $\beta_3$, and
$\beta_3$ in Eq.~$(\ref{49})$ one can easily deduce the transformation
properties of ${\bm\beta}$ and ${\bm\alpha}$:

\be
\alpha_1^\prime=\alpha_1e^{i(2\theta_2^\prime-\theta_1-\theta_2)},\quad
\alpha_2^\prime=\alpha_2e^{i(-2\theta_2^\prime-\theta_1+\theta_2)},
\label{51}
\ee

\be
\beta_1^\prime=\beta_1e^{i(\theta_1^\prime+\theta_2^\prime-2\theta_2)},\quad
\beta_2^\prime=\beta_2e^{i(-\theta_1^\prime+\theta_2^\prime-2\theta_2)},\quad
\beta_3^\prime=\beta_3e^{i(-2\theta_2^\prime-2\theta_2)}.
\label{52}
\ee

\noindent
From these transformation properties it is evident that
$(\alpha_1\alpha_2^*\beta_1^*\beta_2^*\beta_3)$ and hence
$\arg{(\alpha_1\alpha_2^*\beta_1^*\beta_2^*\beta_3)}$ is invariant under
rephasing.

For $n=4$, parametrizing $D(\theta)$ as

\be D(\theta)=
{\rm diag}(e^{i(\theta_1+\theta_2+\theta_3)},
e^{i(-\theta_1+\theta_2+\theta_3)},
e^{-2i\theta_2+i\theta_3},
e^{-3i\theta_3})
\ee

\noindent
and similarly for $D(\theta^\prime)$, one finds that

\bea
V=
A_4(\mbox{\boldmath{$\gamma$}})
A_3(\mbox{\boldmath{$\beta$}})
A_2(\mbox{\boldmath{$\alpha$}})
\rightarrow
V^\prime
&=&
D(\theta^\prime)VD(\theta)\nonumber\\
&=&
A_4(\mbox{\boldmath{$\gamma^\prime$}})
A_3(\mbox{\boldmath{$\beta^\prime$}})
A_2(\mbox{\boldmath{$\alpha^\prime$}})
\nonumber\\
&=&
D(\theta^\prime)
A_4(\mbox{\boldmath{$\gamma$}})
{\rm diag}(e^{i\theta_3},e^{i\theta_3},e^{i\theta_3},e^{-3i\theta_3})
\nonumber\\
&&
A_3(\mbox{\boldmath{$\beta$}})
A_2(\mbox{\boldmath{$\alpha$}})
{\rm diag}(e^{i\theta_1+i\theta_2},e^{-i\theta_1+i\theta_2},e^{-2i\theta_2},1)
\eea

\noindent
The expressions for $\mbox{\boldmath{$\gamma^\prime$}}$ can easily be read off:

\bea
\gamma_1^\prime &=&
\gamma_1e^{i(\theta_1^\prime+\theta_2^\prime+\theta_3^\prime-3\theta_3)},\quad
\gamma_2^\prime =
\gamma_1e^{i(-\theta_1^\prime+\theta_2^\prime+\theta_3^\prime-3\theta_3)},
\nonumber\\
\gamma_3^\prime &=&
\gamma_1e^{i(-2\theta_2^\prime+\theta_3^\prime-3\theta_3)},\quad
\gamma_4^\prime =
\gamma_1e^{-3i(\theta_3^\prime+3\theta_3)}.
\label{55}
\eea

\noindent
A little algebra shows that

\bea
\lefteqn{D(\theta^\prime)A_4(\mbox{\boldmath{$\gamma$}}) {\rm
diag}(e^{i\theta_3},e^{i\theta_3},e^{i\theta_3},e^{-3i\theta_3})
=}\nonumber\\
&& A_4(\mbox{\boldmath{$\gamma$}}^\prime)
{\rm diag}(e^{i(2\theta_2^\prime+2\theta_3^\prime-2\theta_3)},
e^{i(-2\theta_2^\prime+\theta_3^\prime+\theta_3)},e^{i(-3\theta_3^\prime+\theta_3)}
,1),
\eea

\noindent
so that the rest reduces to an $SU(3)$ problem in a $3\times 3$ matrix form

\bea
A_3(\mbox{\boldmath{$\beta$}}^\prime)
A_2(\mbox{\boldmath{$\alpha$}}^\prime)
&=&
{\rm diag}(e^{i(2\theta_2^\prime+2\theta_3^\prime-2\theta_3)},
e^{i(-2\theta_2^\prime+\theta_3^\prime+\theta_3)},
e^{i(-3\theta_3^\prime+\theta_3)},1)\nonumber\\
&& A_3(\mbox{\boldmath{$\beta$}})
A_2(\mbox{\boldmath{$\alpha$}})
{\rm diag}(e^{i\theta_1+i\theta_2},e^{-i\theta_1+i\theta_2},e^{-2i\theta_2}).
\eea

We see that for the $SU(4)$ problem to accompany Eqs.~$(\ref{55})$ we have,

\be
\alpha_1^\prime=\alpha_1e^{i(3\theta_3^\prime-\theta_1-\theta_2-\theta_3)},\quad
\alpha_2^\prime=\alpha_2e^{i(-3\theta_3^\prime-\theta_1+\theta_2+\theta_3)},
\label{58}
\ee

\be
\beta_1^\prime=
\beta_1e^{i(2\theta_2^\prime+2\theta_3^\prime-2\theta_2-2\theta_3)},\quad
\beta_2^\prime=
\beta_2e^{i(-2\theta_2^\prime+\theta_3^\prime-2\theta_2+\theta_3)},\quad
\beta_3^\prime=
\beta_3e^{i(-3\theta_3^\prime-2\theta_2+\theta_3)}.
\label{59}
\ee

With the transformation properties of  $\mbox{\boldmath{$\gamma$}}$,
$\mbox{\boldmath{$\beta$}}$, and $\mbox{\boldmath{$\alpha$}}$ at hand, we can
now systematically construct rephasing invariant quantities out of them as
shown in Ref.~\cite{2}. The three independent invariants turn out to be
$(\alpha_1\alpha_2^*\beta_1^*\beta_2^*\beta_3)$,
$(\beta_2\beta_3^*\gamma_3^*\gamma_4)$, and
$(\beta_1\beta_2^*\gamma_1^*\gamma_2^*\gamma_3)$. The arguments of these
quantities furnish the three independent phase type invariants for the
$SU(3)$ problem. Notice that the first of these is the rephasing invariant for
the $SU(3)$ problem and this is indeed a rather desirable feature of the
recursive parametrization outlined here as one goes from $n$ to $n+1$ one
retains the parameters at the $n^{\rm th}$ level.

In the second form~\cite{3}, for $n=3$, the analogues of Eqs.~(\ref{49}),
(\ref{51}), and (\ref{52}) are

\bea
V&=&
A_3(\mbox{\boldmath{$\beta$}}) A_2(\mbox{\boldmath{$\alpha$}})
\nonumber\\
&=&
\left(
\begin{array}{ccc}
\sigma_1\alpha_2^* & \sigma_1\alpha_1&\beta_1\\
- \beta_1^*\beta_2\alpha_2^*/\sigma_1-\beta_3^*\alpha_1^*/\sigma_1
& -\beta_1^*\beta_2\alpha_1/\sigma_1+\beta_3^*\alpha_2/\sigma_1&\beta_2\\
-\beta_1^*\beta_3\alpha_2^*/\sigma_1+\beta_2^*\alpha_1/\sigma_1&
-\beta_1^*\beta_3\alpha_1/\sigma_1-\beta_2^*\alpha_2/\sigma_1 &\beta_3
\end{array}
\right).
\eea

\be
\alpha_1^\prime=
\alpha_1e^{i(\theta_1^\prime+\theta_2^\prime-\theta_1+\theta_2)},\quad
\alpha_2^\prime=
\alpha_2e^{-i(\theta_1^\prime+\theta_2^\prime+\theta_1+\theta_2)},
\ee

\be
\beta_1^\prime=
\beta_1e^{i(\theta_1^\prime+\theta_2^\prime-2\theta_2)},\quad
\beta_2^\prime=
\beta_2e^{i(-\theta_1^\prime+\theta_2^\prime-2\theta_2)},\quad
\beta_3^\prime=
\beta_3e^{-2i(\theta_2^\prime+\theta_2)},
\ee

\noindent
and the rephasing invariant is $(\alpha_1\alpha_2^*\beta_1^*\beta_2^*\beta_3)$.

For $n=4$ the corresponding equations to $(\ref{55})$, $(\ref{58})$, and
$(\ref{59})$ are

\bea
\gamma_1^\prime&=&
\gamma_1
e^{i(\theta_1^\prime+\theta_2^\prime+\theta_3^\prime-3\theta_3)},\quad
\gamma_2^\prime=
\gamma_1
e^{i(-\theta_1^\prime+\theta_2^\prime+\theta_3^\prime-3\theta_3)},
\nonumber\\
\gamma_3^\prime&=&
\gamma_1
e^{i(-2\theta_2^\prime+\theta_3^\prime-3\theta_3)},\quad
\gamma_4^\prime=
\gamma_1e^{-3i(\theta_3^\prime+3\theta_3)}.
\eea

\be
\alpha_1^\prime=
\alpha_1
e^{i(\theta_1^\prime+\theta_2^\prime+\theta_3^\prime-\theta_1+\theta_2+\theta_3)}
,\quad
\alpha_2^\prime=
\alpha_2
e^{-i(\theta_1^\prime+\theta_2^\prime+\theta_3^\prime+\theta_1+\theta_2+\theta_3)}
.
\ee

\be
\beta_1^\prime=
\beta_1
e^{i(\theta_1^\prime+\theta_2^\prime+\theta_3^\prime-2\theta_2+\theta_3)}
,\quad
\beta_2^\prime=
\beta_2
e^{i(-\theta_1^\prime+\theta_2^\prime+\theta_3^\prime-2\theta_2+\theta_3)}
,\quad
\beta_3^\prime=
\beta_3
e^{-2i(\theta_2^\prime+\theta_3^\prime +\theta_3)}.
\ee

\noindent
The three independent invariants turn out to be
$(\alpha_1\alpha_2^*\beta_1^*\beta_2^*\beta_3)$,
$(\beta_1\beta_2^*\gamma_1^*\gamma_2)$, and
$(\beta_2\beta_3^*\gamma_2^*\gamma_3\gamma_4)$.

\section{\label{seccinco}Constraints due to mod symmetry}

In this section we examine the constraints on the parameters that arise from
demanding that the given $SU(n)$ matrix be mod symmetric, {\it i.e.},
$|V_{ij}|=|V_{ji}|$. For convenience we shall use the first form~\cite{2} for this
discussion.

For $n=3$, mod symmetry requires that $|\alpha_2|=|\beta_2|/\sigma_2$. The
number of independent angle type  invariants comes down from three to two
leaving the phase type invariant unchanged.

For $n=4$, with
$V=A_4(\mbox{\boldmath{$\gamma$}})A_3(\mbox{\boldmath{$\beta$}})
A_2(\mbox{\boldmath{$\alpha$}})$,
after some algebra one finds,

\bea
|V_{14}|&=&
|V_{41}|\Rightarrow
|\alpha_2|=|\gamma_2|/\rho_2,\\
|V_{34}|&=&
|V_{43}|\Rightarrow
|\beta_3|=|\gamma_3|/\rho_3,\\
|V_{23}|&=&
|V_{32}|\Rightarrow
\cos((\delta_1+\delta_2+\delta_3)/2)\times
\nonumber\\
&&\left[\frac{|\beta_2|}{\rho_2}
\cos((\delta_1-\delta_2-\delta_3)/2) +\frac{|\gamma_4|}{\rho_3}\right]=0.
\eea

\noindent
Here $\delta_1$, $\delta_2$, and $\delta_3$ denote the three independent
invariant phases ${\rm arg}(\alpha_1\alpha_2^*\beta_1^*\beta_2^*\beta_3)$,
${\rm arg}(\beta_2\beta_3^*\gamma_3^*\gamma_4)$, and ${\rm
arg}(\beta_1\beta_2^*\gamma_1^*\gamma_2^*\gamma_3)$, respectively. The
equalities $|V_{24}|=|V_{42}|$, $|V_{12}|=|V_{21}|$, and $|V_{13}|=|V_{31}|$
give no new conditions. It can be seen from the above equations that one can
obtain  mod symmetry by requiring

\be
|\alpha_2|=|\gamma_2|/\rho_2,~|\beta_3|=|\gamma_3|/\rho_3,
\quad\delta_1+\delta_2+\delta_3=\pi,
\ee

\noindent
and in this situation the mod symmetric matrix mixing is parametrized
by four angles and two phases.

A simpler moduli symmetric parametrization can be obtained if some of the
eigenvalues $E_i\ (i=1,2,\ldots, n)$, of the $n\times n$ unitary matrix $V$
are equal. For the case when $n-1$ eigenvalues are equal, {\it viz.},
$E_2=E_3=\cdots =E_n$, $V$ can be expressed in terms of $n-1$ real parameters
and only one phase~\cite{4}.

\section{\label{secseis} Comparison with the \lq\lq standard\rq\rq (PDG)
parametrization}

For the case of three generations, the standard or PDG~\cite{pdg08}
parametrization of the mixing matrix is obtained by putting

\be
\alpha_1=c_{12},\quad \alpha_2=s_{12},\quad
\beta_1=s_{13}e^{-i\delta_{13}},\quad \beta_2=s_{23}c_{13},\quad
\beta_3=c_{23}c_{13},
\ee

\noindent
($c_{ij}\equiv\cos\theta_{ij}$ and $s_{ij}\equiv\sin\theta_{ij}$) in
$V=A_3(\mbox{\boldmath{$\beta)$}}A_2(\mbox{\boldmath{$\alpha$)}}$ with
$A_3(\mbox{\boldmath{$\beta)$}}$ and $A_2(\mbox{\boldmath{$\alpha$)}}$ given by
Eqs.~(\ref{36}) and (\ref{35}), respectively.

The extension of the mixing matrix to four generations is given by
$V=A_4(\mbox{\boldmath{$\gamma
$}})A_3(\mbox{\boldmath{$\beta$}})A_2(\mbox{\boldmath{$\alpha $}})$, where
$A_4(\mbox{\boldmath{$\gamma $}})$ is given by Eq.~$(\ref{37})$ with
$\mbox{\boldmath{$\beta $}}$ and $\mbox{\boldmath{$\alpha $}}$ as before and

\begin{equation}
\gamma_1=s_{14}e^{-\delta_{14}},\quad
\gamma_2=c_{14}s_{24}e^{-i\delta_{24}},\quad
\gamma_3=s_{34}c_{24}c_{14},\quad\gamma_4=c_{34}c_{24}c_{14},
\end{equation}

\noindent
which conveniently reduces to the case of three generations when
$\theta_{14}$, $\theta_{24}$, and $\theta_{34}$ are all set equal to zero.

We note here that the parametrization given above is closely related to the
Harari-Leurer parametrization~\cite{5} where the mixing matrix is expressed as
an ordered  product of essentially $2\times 2$ \lq\lq rotation\rq\rq  matrices.
Our parametrization results when one suitably combines the factors appearing in
that form. For instance, in the $4\times 4$ case, the Harari-Leurer form for the
mixing matrix has the structure
$\Omega_{34}\Omega_{24}\Omega_{14}\Omega_{23}\Omega_{13}\Omega_{12}$ and reduces
to our form by the identifications $A_4(\mbox{\boldmath{$\gamma $}})\equiv
\Omega_{34}\Omega_{24}\Omega_{14}$,
$A_3(\mbox{\boldmath{$\beta$}})=\Omega_{23}\Omega_{13}$, and
$A_2(\mbox{\boldmath{$\alpha $}})=\Omega_{12}$, provided we choose $\gamma_4$,
$\beta_3$, and $\alpha_2$ to be real.

\section{\label{conclusions} Conclusions}

In this work we have examined in detail the question of
parametrizing quark flavor mixing matrices for three and four
flavors within the framework of the recursive parametrization
developed in Refs.~\cite{2} and \cite{3}. In particular we have shown,
given the matrix, how to determine the corresponding parameters.
We have also studied in detail aspects of rephasing invariants in
this parametrization scheme and have derived  conditions for the
mixing matrix to be moduli symmetric.

\begin{acknowledgments}

VG and G.~S-C would like to thank CONACyT (M\'exico) for partial
support.

\end{acknowledgments}

\end{document}